\documentclass{elsarticle}

\usepackage{amssymb}
\usepackage{amsmath}
\usepackage{hyperref}

\usepackage{listings,xcolor}
\journal{Physica A}

\begin{document}

\begin{frontmatter}

\title{Correlation between the Hurst exponent and the maximal Lyapunov exponent: examining some low-dimensional conservative maps}

\author{Mariusz Tarnopolski}
\address{Astronomical Observatory, Jagiellonian University, Orla 171, PL-30-244 Krak\'ow, Poland}
\ead{mariusz.tarnopolski@uj.edu.pl}

\begin{abstract}
The Chirikov standard map and the 2D Froeschl\'e map are investigated. A few thousand values of the Hurst exponent (HE) and the maximal Lyapunov exponent (mLE) are plotted in a mixed space of the nonlinear parameter versus the initial condition. Both characteristic exponents reveal remarkably similar structures in this space. A tight correlation between the HEs and mLEs is found, with the Spearman rank $\rho=0.83$ and $\rho=0.75$ for the Chirikov and 2D Froeschl\'e maps, respectively. Based on this relation, a machine learning (ML) procedure, using the nearest neighbor algorithm, is performed to reproduce the HE distribution based on the mLE distribution alone. A few thousand HE and mLE values from the mixed spaces were used for training, and then using $2-2.4\times 10^5$ mLEs, the HEs were retrieved. The ML procedure allowed to reproduce the structure of the mixed spaces in great detail.
\end{abstract}


\end{frontmatter}

\section{Introduction}
\label{sec1:intr}

Dynamical systems play a crucial role in the description of the physical reality, being applied in fields such as cosmology \cite{Wainwright}, astrophysics \cite{Manos,Zotos}, nuclear physics \cite{MacKay}, environmental science \cite{Wang}, financial analysis \cite{Gao}, among others. In particular, nonlinear systems often exhibit chaotic behavior \cite{Alligood,Ott}, e.g. Chirikov standard map \cite{Chirikov,Lichtenberg} being a discrete volume preserving 2D example, or Lorenz \cite{Lorenz} and H\'enon-Heiles \cite{Henon&Heiles} systems, being 3D dissipative and 4D conservative continuous systems, respectively. Continuously, new chaotic systems are being discovered \cite{Alpar,Zhang}. Conservative systems, being Hamiltonian \cite{Bountis&Skokos,Lowenstein}, exhibit a complicated mixture of chaotic and regular components in the phase space and do not posses a strange attractor \cite{Greiner}. Moreover, a parameter--initial condition mixed space allowed to properly trace the route to chaos via period doubling in the Chirikov map \cite{Manchein}. On the other hand, a question about inferring chaotic dynamics from a scalar time series was also raised and efficiently answered decades ago \cite{Hegger,Rosenstein,Wolf}.

Time series may be, in general, described by their statistical properties. One of its descriptors is the Hurst exponent (HE) \cite{Hurst,Mandelbrot,Mandelbrot&Wallis}, which is a measure of persistency, or long-range memory (or lack of thereof) \cite{Machado} that is widely used, e.g., in financial analyses \cite{Carbone2} and Solar physics \cite{Suyal}. It proved to be a useful indicator of morphological type in astrophysical processes, too \cite{MacLachlan,Tarnopolski}. An HE, denoted $H$, related to persistent (long-term memory) processes is greater than $1/2$, while anti-persistent (short-term memory) ones yield $H<1/2$.

\subsection{Motivation}\label{motiv}

The research presented in this paper is inspired by the findings in Ref. \cite{Suyal}, where a time series analysis of the sunspot number was performed. The authors, for illustrative purposes, examined also a chaotic time series generated from the celebrated Lorenz equations, which resulted in an $H>1/2$ (obtained with the R/S algorithm \cite{Mandelbrot&Wallis}), indicating persistent behavior. However, it was shown recently \cite{Araujo} that the decay of correlations for the Lorenz system is exponential, therefore $H=1/2$. In case of the Chirikov standard map, the same exponential decay was observed at the border of chaos \cite{Chirikov2}. It is not that obvious that $H$ must always be around $1/2$, especially when motion deep in the chaotic zone---not only at the border---is considered. Also, the time series from the Lorenz and Chirikov systems are visually quite different. In a long run, those from the Lorenz system resemble a white noise. On the other hand, the time series of the Chirikov map more resemble those of a fractional Brownian motion \cite{Tarnopolski2}. Therefore, the correlations between the maximal Lyapunov exponents (mLEs) and $H$ are to be examined herein.

The Chirikov standard map \cite{Chirikov,Lichtenberg,Meiss} and the two-dimensional Froeschl\'e map \cite{Froeschle1971} are chosen to work with due to the fact that they are well examined and simple in their formulation. In particular, (i) Ref. \cite{Manchein} provides an immediate comparison with the results obtained herein for the Chirikov standard map, and (ii) the 2D Froeschl\'e map has been also widely examined in the literature \cite{Froeschle3,Skokos1,Froeschle1971}. Also, for the former reason, the mLE is employed as a chaos indicator.

%{\bf For completeness, it should be noted that the goal of the presented research could be also achieved with other well-established indicators, such as the fast Lyapunov indicator (FLI) \cite{Froeschle1,Froeschle2,Froeschle3}, mean exponential growth factor of nearby orbits (MEGNO) \cite{Cincotta1,Cincotta2,Maffione} (note that there is a precise analytical relation between the FLI and MEGNO \cite{Mestre}, due to which these two indicators are not independent), smaller alignment index (SALI) \cite{Bountis&Skokos,Bountis,Manos1,Skokos1}, the generalized alignment index (GALI) \cite{Bountis&Skokos,Manos2,Manos3,Skokos2}, the relative Lyapunov indicator (RLI) \cite{Sandor1,Sandor2}, or the rotation index \cite{Voglis}, to mention only a few that rely on the solution of the variational equation (hence these methods are frequently called variational ones). Alternatively, there are diagnostics based on frequency decomposition (the frequency map analysis), in which the frequency vectors for various initial conditions form the frequency map, which is analysed by studying its regularity \cite{Laskar1,Laskar2,Laskar3}. This method is conceptually different from the variational indicators as it does not employ the concept of divergence of nearby orbits. }

\subsection{Aims and structure}

In this work, correlations between mLEs and $H$ are investigated for the Chirikov and 2D Froeschl\'e maps. Machine learning (ML) is performed to successfully reproduce the statistical $H$ distribution given an mLE distribution, showing that the connection between these two characteristic exponents allows to infer the $H$ values based on the mLEs only.

This paper is organized in the following manner. In Sect.~\ref{sec2:meth}, the mLE and $H$ are briefly characterized, and the algorithms for their computation are outlined. In Sect.~\ref{sec3:model}, the maps under consideration are defined, and the performance of the mLE and $H$ extraction algorithms is demonstrated. The main results are presented in Sect.~\ref{sec4:res}, which is followed by the ML approach in Sect.~\ref{sec5:mach}. Discussion and concluding remarks are gathered in Sect.~\ref{sec6:disc}. A \textsc{Mathematica} computer algebra system is used throughout.

\section{Methods}
\label{sec2:meth}

%The algorithms for the mLE and $H$ computation, described below, were implemented in \textsc{mathematica}\textsuperscript{\textregistered}, and are available on request.

\subsection{Maximal Lyapunov exponent}\label{mle}

The Lapunov exponent $\lambda$ (LE) \cite{Ott,Bountis&Skokos,Lowenstein,Wolf,Baker,Benettin1,Benettin2,Oseledec,Skokos,Tabor} is a measure of the mean exponential divergence (or convergence) of two initially nearby orbits of a dynamical system in its phase space in a time limit of infinity. The maximal LE (mLE), $\lambda_1$, indicates chaos if $\lambda_1>0$. For conservative maps with $N=2$, the relation $\lambda_1+\lambda_2=0$ holds, and hence knowing the mLE, the other LE is immediately known also \cite{Ott,Bountis&Skokos,Tabor}.

Consider a map $M$ acting on an $N$-dimensional phase space vector $\mathbf{y}$,
\begin{equation}
\mathbf{y}_{n+1}=M(\mathbf{y}_n),
\label{eq1}
\end{equation}
with $n=0,1,2,\ldots$, an initial condition $\mathbf{y}_0$, and an infinitesimal deviation vector $\mathbf{w}$, evolving with each iteration according to the variational equation
\begin{equation}
\mathbf{w}_{n+1}={\rm D}M(\mathbf{y}_n)\cdot \mathbf{w}_n,
\label{eq2}
\end{equation}
where ${\rm D}M(\mathbf{y}_n)$ is the Jacobian matrix of the map $M$ evaluated at $\mathbf{y}_n$. It follows from Eq.~(\ref{eq2}) that $\mathbf{w}_n={\rm D}M^n(\mathbf{y}_0)\cdot \mathbf{w}_0$, where
\begin{equation}
{\rm D}M^n(\mathbf{y}_0)={\rm D}M(\mathbf{y}_{n-1})\cdot {\rm D}M(\mathbf{y}_{n-2})\cdot\ldots\cdot {\rm D}M(\mathbf{y}_0).
\label{eq3}
\end{equation}
Then, taking without loss of generality $||\mathbf{w}_0^i||=1$, the LEs are given by \cite{Ott,Tabor}
\begin{equation}
\lambda_i=\lim\limits_{n\rightarrow\infty}\frac{1}{n}\ln||\mathbf{w}^i_n||\simeq\frac{1}{2n}\ln|h_i|,
\label{eq4}
\end{equation}
where $i=1,\ldots,N$ corresponds to the eigenvalues $h_i$ of the matrix $H_n(\mathbf{y}_0)=\left[{\rm D}M^n(\mathbf{y}_0)\right]^{\intercal}{\rm D}M^n(\mathbf{y}_0)$. In practical implementations, the limit $n\rightarrow\infty$ is replaced by $n$ sufficiently large, leading to the finite time LEs (FTLEs), usually being valid approximations of the LEs.

\subsection{Hurst exponent}\label{he}

The quantity $H$, introduced by H. E. Hurst in 1951 to model statistically the cycle of Nile floods \cite{Hurst,Mandelbrot}, is a measure of long-term memory of a process. A persistent process has long-term memory, and as such is characterized by $H>1/2$. The $H$ value can be smaller than $1/2$; the process is then called anti-persistent, and it posseses short-term memory. The HE is also related to the autocorrelation of a process, i.e. to the rate of its decrease with increasing lag.\footnote{Via a power law, hence the notion of an {\it exponent}; see \cite{Tarnopolski2}.} Finally, $H$ is bounded to the interval $(0,1)$. Its properties can be summarized as follows \cite{Tarnopolski2}:
\begin{enumerate}
\item $0<H<1$,
\item $H=1/2$ for a white noise (uncorrelated process),
\item $H>1/2$ for a persistent (long-term memory, correlated) process,
\item $H<1/2$ for an anti-persistent (short-term memory, anti-correlated) process.
\end{enumerate}

Among many existing computational algorithms for the estimation of $H$ \cite{Mandelbrot&Wallis,DFA1,DFA2,Jones,Simonsen}, detrended moving average (DMA) \cite{Carbone} is used herein due to its simplicity and closed-form treatment \cite{Arianos}. In this method, first a moving average $\widetilde{y}_w(i)$ of a time series $y(i)$, being a realization of a process under consideration, with equally spaced points $i=1,2,\ldots,N_{\rm max}$, is computed:
\begin{equation}
\widetilde{y}_w(i)=\frac{1}{w}\sum\limits_{k=0}^{w-1}y(i-k),
\label{eq6}
\end{equation}
i.e. the average of $y$ for the last $w$ data points, where $w\in[w_{\rm min},w_{\rm max}]$ is the sample window, with a step of $\Delta w$. The moving average captures the trend of the signal over a (discretized) time interval of length $w$ \cite{Vandewalle}. Next, the variance of $y(i)$ with respect to $\widetilde{y}_w(i)$ is defined by
\begin{equation}
\sigma^2_{MA}=\frac{1}{w_{\rm max}-w}\sum\limits^{w_{\rm max}}_{i=w}\left[y(i)-\widetilde{y}_w(i)\right]^2,
\label{eq7}
\end{equation}
where $w_{\rm max}\ll N_{\rm max}$. As it obeys the power law $\sigma_{MA}\propto w^H$, $H$ is obtained as a slope of a linear regression in the $\ln \sigma_{MA}\!-\!\ln w$ plane.

\section{Models}
\label{sec3:model}

Two common conservative 2D maps are considered: the Chirikov standard map \cite{Chirikov} and the 2D Froeschl\'e map \cite{Froeschle1971}. Both are symplectic, governed by a single nonlinear parameter, and they exhibit, besides strictly regular and chaotic, also sticky behavior.

Lacking an entrenched name, to differentiate from its more popular 4-dimensional version, often termed simply the Froeschl\'e map, its 2-dimensional version is herein named the {\it 2D Froeschl\'e map} as it first appeared in Ref. \cite{Froeschle1971}. Note there are some naming ambiguities in the literature, e.g. in \cite{Howard} the Chirikov standard map is called a Froeschl\'e map.

\subsection{Chirikov map}

\begin{figure}
\centering
\includegraphics[width=\columnwidth]{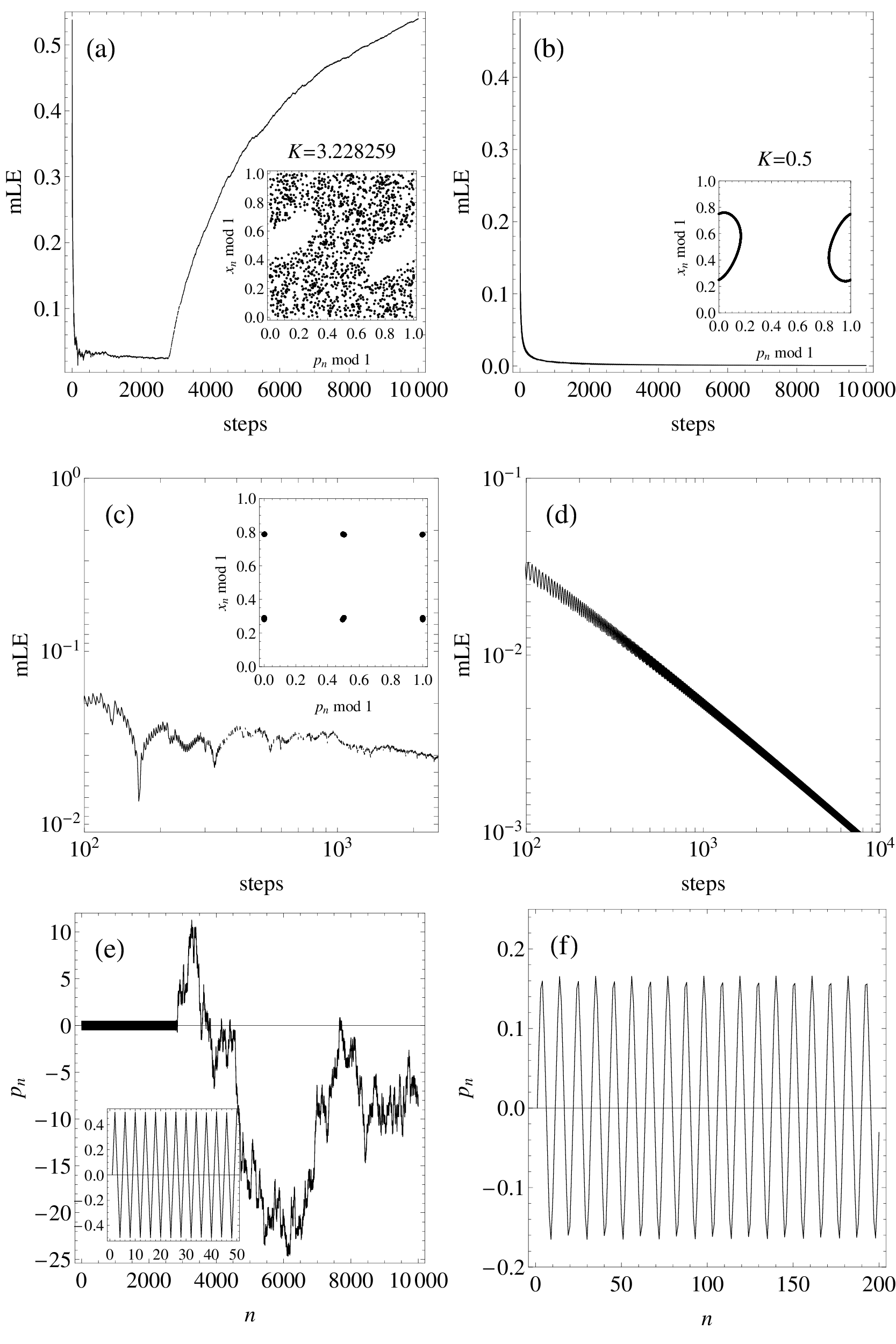}
\caption{Convergence plots of the mLE for (a) chaotic (which exhibits also sticky behavior), with $(p_0,x_0)=(0, 0.2865)$, and (b) regular orbit, with $(p_0,x_0)=(0, 0.25)$. The insets show phase space portraits of the orbits. The map was iterated for $10^4$ steps. (c) Log-log plot of the mLE convergence for the chaotic orbit in the region of sticky motion, i.e. up to 2500 steps. The inset shows the corresponding phase space portrait. (d) Linear decline in case of the regular orbit. Note different scales in (c) and (d). In (e) the time series of $p_n$ is displayed for the chaotic orbit; the inset shows its evolution for the first 50 iterations in the sticky region; (f) shows part of a $p_n$ time series corresponding to a regular orbit.}
\label{figB}
\end{figure}

The Chirikov standard map \cite{Chirikov,Lichtenberg,Meiss} in the form
\begin{eqnarray}
\left\{ \begin{array}{ll}
p_{n+1}=p_n+\frac{K}{2\pi}\sin(2\pi x_n), \\
x_{n+1}=x_n+p_{n+1}, \\
\end{array} \right.
\label{eq8}
\end{eqnarray}
is examined. It is conservative and governed by a single nonlinear parameter $K$. Global chaos occurs for $K>K_c\simeq 0.97163540631$ \cite{Greene,MacKay2}. This map can also exhibit, besides strictly regular and chaotic, also sticky behavior \cite{Afraimovich}, which means that the orbit may look quite regular for some time and only after a sufficiently large number of iterations $n$ its chaotic features start to be clearly visible (a transition from temporarily regular behavior to apparently chaotic variations at some $n_0$, i.e. the orbit might be easily missclasified if $N_{\rm max}\simeq n_0$). An example of such situation is shown in Fig.~\ref{figB}. The mLE seems to slowly decrease to zero at first [convergence plots in Fig.~\ref{figB} (a) and (c)], but at $n\approx n_0 = 2500$ it suddenly starts to increase and plateaus at a positive nonzero value. The stickiness is also clearly visible in the time series of $p_n$ [Fig.~\ref{figB} (e)], which looks regular at first, but then starts to oscillate roughly. Contrary to this, the convergence plot of a purely regular orbit [Fig.~\ref{figB} (b)] exhibits an $n^{-1}$ decrease, clearly visible in a log-log plot [Fig.~\ref{figB} (d)] as a straight line, indicating that $\lambda_1\rightarrow 0$ when $n\rightarrow\infty$. The time series of $p_n$ is also completely regular for the whole range of $n$ that it was iterated in [only a part is displayed in Fig.~\ref{figB} (f) for the sake of clarity].
It was verified for several randomly chosen initial conditions and values of $K$ that collecting the mLE after $N_{\rm max}=10^4$ iterations is sufficient for the purpose of this work, and hence is employed hereinafter. While there might happen orbits that exhibit stickiness for a longer time, even greater than $10^4$ iterations, or that are initially chaotic but then stick near a resonant island for some time, leading to false classifications, these instances are not frequent enough to obscure the overall statistical picture revealed herein. An analysis of outliers is hence out of scope of the current research.

In the estimation of $H$ it was chosen to discard the first 4000 iterations due to a possibility of encountering sticky behavior, i.e. for the DMA algorithm a $p_n$ time series of total length 6000, with $n\in(4000,10\,000]$, $n\in\mathbb{N}$, is used; $p_n$ is employed as $x_n$ is monotonic, hence yielding $H=1$ \cite{Katsev}.

Having a time series of $p_n$, to obtain $H$ according to Eq.~(\ref{eq6}) and (\ref{eq7}), the following parameters are fixed and used throughout this paper: $w_{\rm min}=10$, $w_{\rm max}=300$, $\Delta w=10$. Next, a linear regression is performed on the $\ln\sigma_{MA}\!-\!\ln w$ relation, and the slope of the fit is the estimated $H$ value. To confirm that the fitting is performed correctly, i.e. a line is fitted in the linear part of the plot with sufficient accuracy, a number of statistical indicators are computed. First, the standard error of the slope is retrieved. Next, the end points $(a,b)$ of the 99\% confidence interval of the slope are used to calculate its width, $\alpha:=b-a$. Finally, the Pearson coefficient $R^2$ is obtained. The standard error and $\alpha$ should be small compared to the corresponding $H$ value, and $R^2$ should be close to unity to conclude that the fitting was reliable.

\begin{figure}
\centering
\includegraphics[width=0.75\columnwidth]{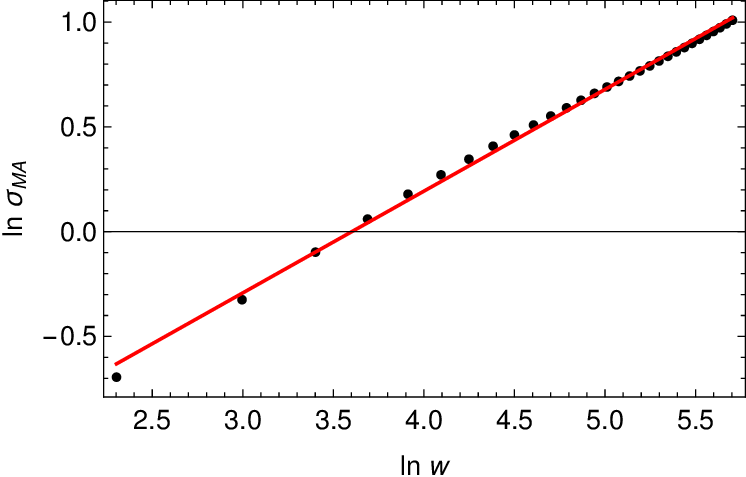}
\caption{An example of $H$ estimation for the time series from Fig.~\ref{figB} (e), after discarding the first 4000 iterations. See text for details.}
\label{figDMA}
\end{figure}

An example of such an approach is illustrated in Fig.~\ref{figDMA}, in which case the results of the linear regression are as follows: $H=0.4856$, the standard error is equal to 0.0047, $\alpha=0.4985-0.4727=0.0258$, and $R^2=0.999$. This convincingly places the $H$ estimate slightly below the value of 0.5 for this time series.
It might happen, however, that some $H$ exceeds unity, which is a meaningless result based on the mathematical theory (see Sect.~\ref{he}), but is not surprising in numerical computations. Fortunately, it will turn out that these values are a negligible fraction in the statistic, and will not affect the main results and conclusions.

\subsection{2D Froeschl\'e map}

The 2D Froeschl\'e map \cite{Froeschle1971} in the form
\begin{eqnarray}
\left\{ \begin{array}{ll}
p_{n+1}=p_n-k\sin(x_n+p_n), \\
x_{n+1}=x_n+p_n, \\
\end{array} \right.
\label{eq9}
\end{eqnarray}
is examined. It is formulated similarly to the map in Eq.~(\ref{eq8}), and belongs to the same family. This map may also exhibit regular and chaotic motion; in particular, sticky behavior may be observed. Again, it was verified that $10^4$ iterations are sufficient for the computation of the mLE, as it was for the Chirikov standard map, and in order to obtain the $H$ estimates, the first 4000 steps are discarded from the $p_n$ time series ($x_n$ is again monotonic), and its remaining part (of length 6000) is the input for the DMA algorithm.

While it might happen for some orbits (in case of both the Chirikov and 2D Froeschl\'e maps) that the total number of iterations $N_{\rm max}$, or the initial number of iterations discarded, will not be sufficient to skip over transient behavior, e.g. stickiness (moreover, a chaotic trajectory can become sticky after very long times; in area-preserving maps, such as the Chirikov and 2D Froeschl\'e maps, stickiness generically occurs at the border of 2-dimensional Kolmogorov-Arnold-Moser (KAM) island \cite{Lichtenberg,Kantz,Silva}), this should not be a concern, as several thousands of orbits are to be examined, and even if a small fraction will be missclasified, it should not affect the generality of the final results, which are of statistical character.

For completeness, it should be pointed out that the computation of $H$ using the DMA method was about 18 times more time-consuming than of the mLE.

\section{Results}
\label{sec4:res}

\subsection{Chirikov standard map}
\label{subsec31:res}

\begin{figure}%[h!]
\centering
\includegraphics[width=\columnwidth]{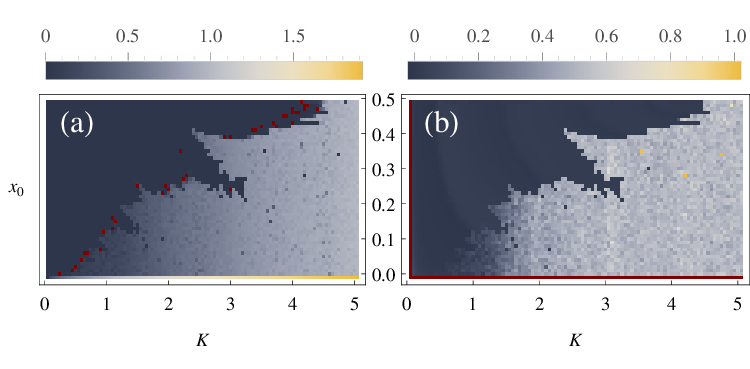}
\caption{(a) mLEs and (b) $H$ in the mixed space $K\times x_0$ with $p_0=0$ for Chirikov standard map, and a grid of $101\times 51=5151$ points. Red points mark indeterminate values. Note different color scales used.}
\label{figur1}
\end{figure}

A map of mLEs in a mixed space $K\times x_0$ on a grid of $101\times 51$ points with $p_0=0$ is shown in Fig.~\ref{figur1}(a). This is a recalculation of Fig.~2 in Ref. \cite{Manchein} but due to symmetry only non-negative $x_0$'s are displayed. Although the overall picture sketched by both Figures (i.e., Fig.~\ref{figur1}(a) herein and Fig.~2 in Ref. \cite{Manchein}) is consistent with each other, note slight differences in mLE values: according to Ref. \cite{Manchein}, the biggest mLE attained in the $K\times x_0$ space is $\approx 1.2$, while herein the mLEs reach a value of $\approx 1.92$. The difference is most likely caused by applying different algorithms: in Ref. \cite{Manchein} an mLE was extracted from a time series as described by Wolf et al. \cite{Wolf} (C.~Manchein, personal communication), while herein the mLEs were computed according to the method described in Sect.~\ref{mle}. The difference is especially striking for the unstable line $x_0=0$, which is an exceptional line and it appears that the Wolf et al. method \cite{Wolf} could not fully grasp the underlying dynamics along this line. It was verified with some randomly chosen points on this line, as well as for the point from Fig.~\ref{figB}(a), that the final value of the mLE varies by a factor of $\sim 6$ when different, yet reasonable, input parameters for the Wolf et al. algorithm \cite{Wolf} are used. The method employed herein, described in Sect.~\ref{mle}, is free of such numerous parameters.

Note also that in Fig.~\ref{figur1}(a) some mLEs were undetermined (red points), because the eigenvalues $h_i$ of the matrix $H_n(\mathbf{y}_0)$ are numerical zeros in some points, and the mLE is given according to Eq.~(\ref{eq4}) as their logarithm. Judging from the analysis in Ref. \cite{Manchein}, the verification of the method performed in Sect.~\ref{sec3:model}, and the relative difference between neighboring (in the mixed space) mLEs, the division between regular and chaotic region is quite sharp. Because the mLEs are in fact FTLEs, they will in fact never converge to zero, and due to different convergence rates of different regular orbits, small-valued mLEs ($\sim 0.01-0.02$ in this case) are effectively ascribed to the regular domain. As this work is not focused on detailed aspects of the mixed space, but rather on general features---in particular, a rough differentiation between {\it low} and {\it high} mLEs is sufficient for the statistical analysis performed further on---the rich properties of the regular domain in the mixed space will not be discussed hereinafter.\footnote{The reader is referred to Ref. \cite{Manchein} for details regarding the interpretation of FTLEs in this context.}

Next, a similar plot for $H$ was drawn and is displayed in Fig.~\ref{figur1}(b). The unstable line $x_0=0$ had to be discarded as the $H$ values were undetermined by the algorithm. This is because, according to Eq.~(\ref{eq8}), the initial point $(0,0)$ is mapped also onto $(0,0)$ for all $K$, so the time series of $p_n$ is constant, hence the DMA method leads to $\sigma_{MA}=0$. If this is fed to the automated procedure of fitting a straight line in a log--log plot for the $H$ extraction described in Sect.~\ref{sec3:model}, it results in indeterminate values. The same situation occurs for $K=0$ and varying $x_0$.

\begin{figure}
\centering
\includegraphics[width=\columnwidth]{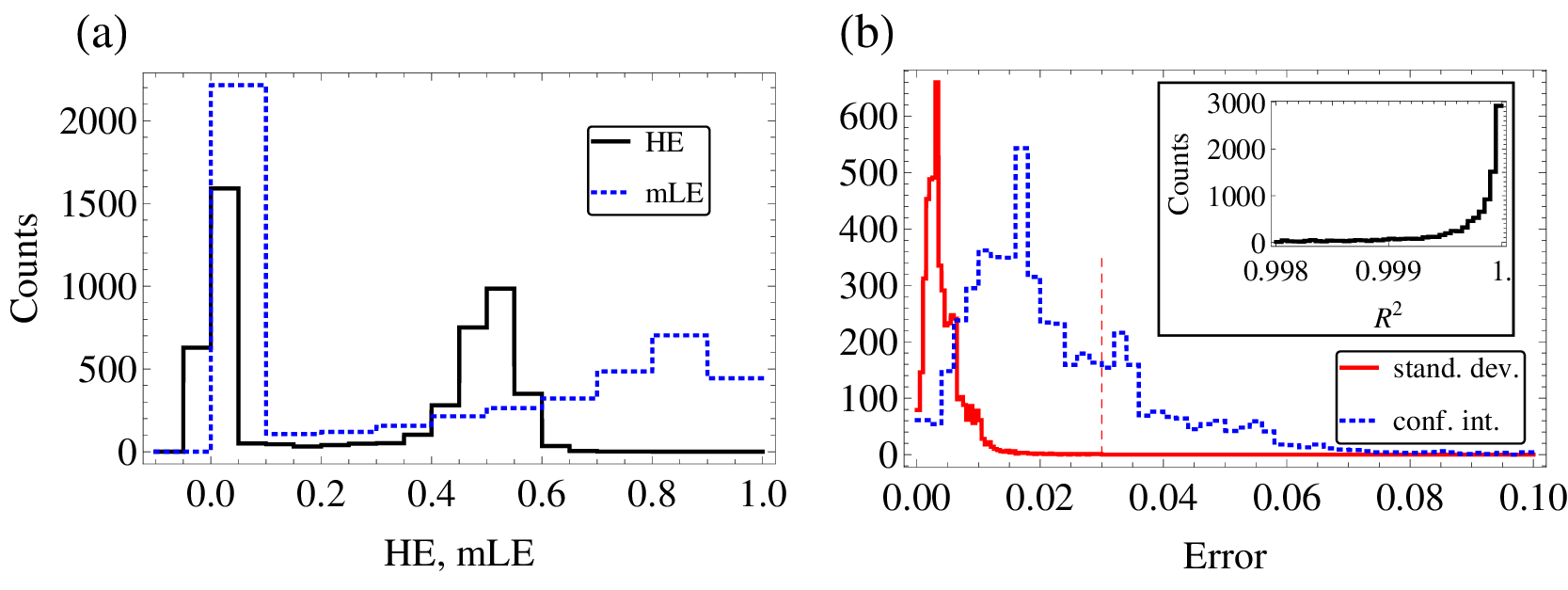}
\caption{(a) Distributions of $H$ and mLE for the Chirikov standard map. Two distinct peaks are related to regular and chaotic domain of the mixed space. The mLE values greater than 1.0 are not displayed for the sake of clarity as they account for only 1.6\% of all values, and are mainly located along the unstable line $x_0=0$. (b) $H$ estimates' errors; solid red -- standard deviation, std, of the fitted slope, dashed blue -- width of the 99\% confidence interval, $\alpha$. Maximal value of std is 0.03 (marked with a vertical red line), while $\alpha$ is not greater than 0.164. Inset shows the Pearson coefficient $R^2$; displayed bins contain 94\% of counts. The minimal $R^2$ is 0.9732.}
\label{figur2}
\end{figure}

In Fig.~\ref{figur2}(a), the histograms of mLEs and $H$ are displayed. The height of the peak around small mLE values is related to the size of the regular region in the mixed space, and the mLEs related to chaotic motion have a peak at $\sim 0.8-0.9$. In the $H$ distribution, there are two prominent peaks: one at small values, $H\sim 0$, and one located at $H$ slightly exceeding 0.5. Comparing this distribution with Fig.~\ref{figur1}(b), one may associate the two peaks with regular and chaotic domains of the mixed space, respectively.

Histograms in Fig.~\ref{figur2}(b)---the standard deviation, range $\alpha$ of the 99\% confidence interval, and Pearson's $R^2$ of each $H$ estimate---convince that the fitting procedure returned reliable estimates of $H$: the standard deviation of the slope does not exceed 0.03, and the Pearson $R^2$ does not fall below 0.9732. The range $\alpha$ of the confidence interval reaches a value as high as 0.164, but it has a mode of 0.02. Overall, the extracted $H$ valuess appear to be reasonable estimates.

\begin{figure}
\centering
\includegraphics[width=0.75\columnwidth]{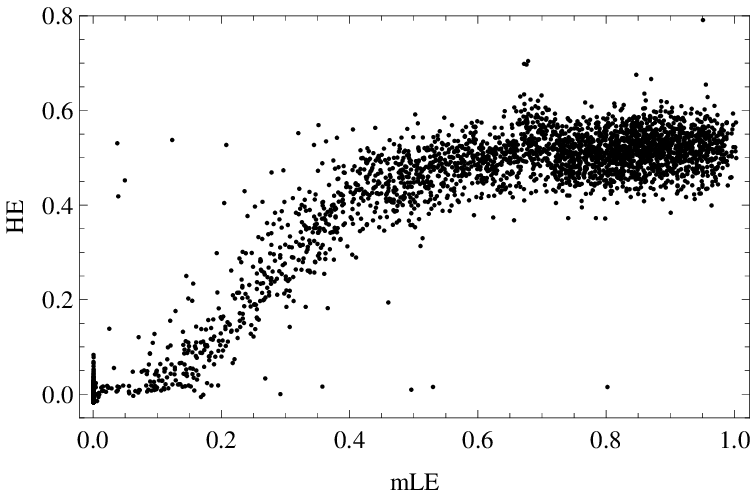}
\caption{Scatter plot for mLE--HE relation in case of the Chirikov standard map from Eq.~(\ref{eq8}); $N_p=4962$ points are displayed for which $\rho=0.83$ and its $p$-value is numerically equal to zero.}
\label{figur3}
\end{figure}

After discarding outcomes being indeterminate for either mLE or $H$, $N_p=4962$ points with numerical values were left for which a scatter plot is shown in Fig.~\ref{figur3}. The correlation is nonlinear, hence a Spearman rank $\rho$ is used to quantify its strength, and yielded $\rho=0.83$, with $p$-values numerically equal to zero. Hence, the mLE--HE relation is a tight one.

Despite a negligible amount of outliers, the correlation between mLE and $H$ is very high, and the computation of $H$ took $\sim 18$ times longer than of mLEs, this relation will provide a useful insight into the statistical distribution of $H$ based on easier and faster to compute mLEs.

\subsection{2D Froeschl\'e map}
\label{subsec32:res}

\begin{figure}
\includegraphics[width=\columnwidth]{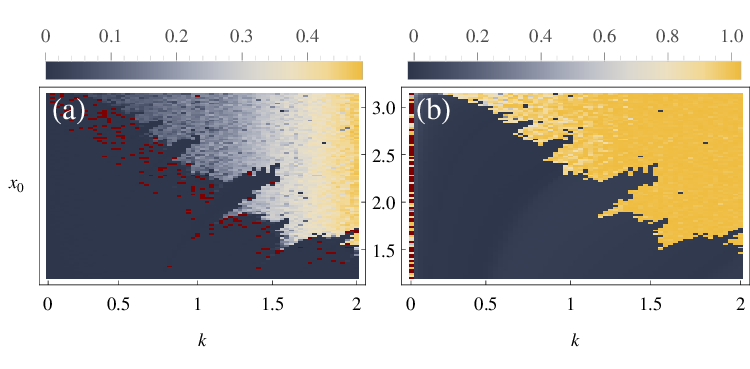}
\caption{(a) mLEs and (b) $H$ in the mixed space $k\times x_0$ with $p_0=0$ for a 2D Froeschl\'e map, and a grid of $67\times 98=6566$ points. Red points mark indeterminate values. Note different color scales used.}
\label{figur4}
\end{figure}

In order to check the robustness of the results obtained for the Chirikov standard map, the 2D Froeschl\'e map given by Eq.~(\ref{eq9}) is investigated in the same manner as in the previous subsection. Fig.~\ref{figur4}(a) and (b) show the mLE and $H$ distributions, respectively, in the mixed space $k\times x_0$ with $p_0=0$, and exhibit the same structure as the Chirikov standard map does, i.e. bifurcating tongues of regular motion creeping in the chaotic zone. A significant difference is that the values of $H$ gather around two extreme values, i.e. $\sim 0$ for regular and $\sim 1$ for chaotic regions, different than they were for the Chirikov standard map, and with a negligible gradient in between.

\begin{figure}%[h!]
\includegraphics[width=\columnwidth]{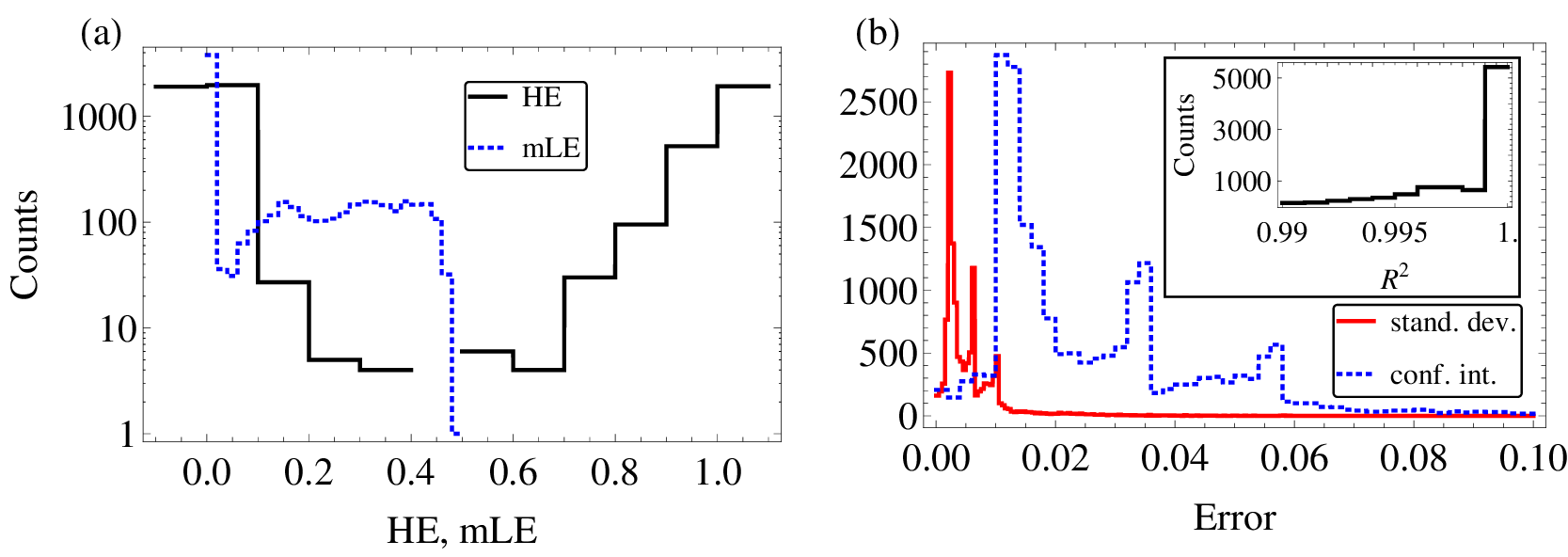}
\caption{(a) Distributions of $H$ and mLE for the 2D Froeschl\'e map. Two distinct $H$ peaks are related to regular and chaotic domain of the mixed space. The mLE values are not greater than 0.5 and have an approximately uniform distribution inside the chaotic region. Note this is a semi-log plot. (b) $H$ estimates' errors; solid red -- standard deviation, std, of the fitted slope, dashed blue -- width of the 99\% confidence interval, $\alpha$. Maximal value of std is 0.19, while $\alpha$ spans an interval from 0 to 1.04. Inset shows the Pearson coefficient $R^2$; displayed bins contain 84\% of counts. The minimal $R^2$ is 0. Extreme values of $\alpha$ and $R^2$ are rare enough to be treated as outliers.}
\label{figur5}
\end{figure}

Statistical distributions, displayed in Fig.~\ref{figur5}(a) (note a logarithmic scale for the ordinate), reveal an almost uniform distribution of mLEs for chaotic motion. Fittings performed for the estimation of $H$ are not as reliable as in the former case, though. For example, the Pearson coefficient $R^2$ can attain a value as small as 0. On the other hand, a majority of fittings are characterized by errors small enough to allow examining at least general features of the $H$ distribution [compare with Fig.~\ref{figur5}(b)]. Although the maximal 99\% confidence interval range can be as wide as the whole theoretical range of $H$ values (spanning a unit interval from 0 to 1), the standard deviations do not exceed the value of 0.19, and (i) are centered around small absolute values, as indicated by the histogram in Fig.~\ref{figur5}(b), and (ii) allow to distinguish chaotic from regular motions due to clearly separated peaks near extremal values of $H$.

\begin{figure}%[h!]
\begin{center}
\includegraphics[width=0.75\columnwidth]{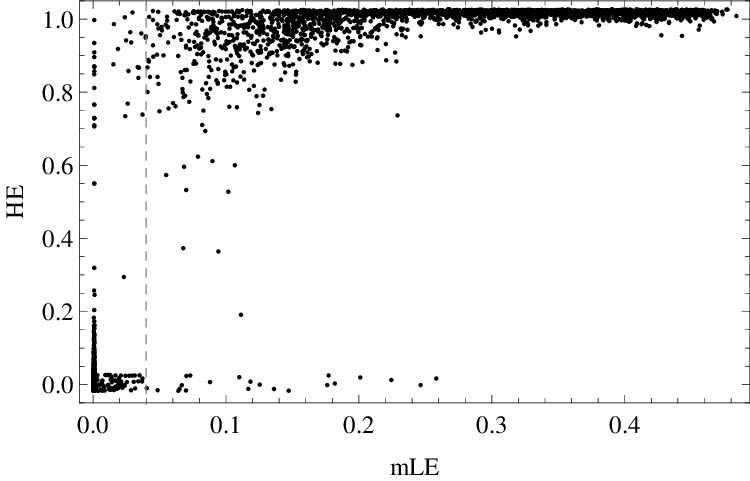}
\caption{Scatter plot for mLE--HE relation in case of the 2D Froeschl\'e map from Eq.~(\ref{eq9}); $N_p=6314$ points are displayed for which $\rho=0.75$ and its $p$-value is numerically equal to zero.}
\label{figur6}
\end{center}
\end{figure}

Thus, as the scatter plot in Fig.~\ref{figur6} for this map does not posses as unambiguous structure as for the previous one, the distributions in Fig.~\ref{figur5} assure that the relation between the mLE and $H$ is correctly grasped, at least at a first approximation. Note that the flat cut-off from above in Fig.~\ref{figur6} is a true feature and is not an artifact due to dropping indeterminate values (as $0<H<1$). Moreover, the higher the mLE, the smaller the scatter among the corresponding $H$, and the small $H$ values are separated accurately from larger ones (or---equivalently in this case---values corresponding to regular and chaotic zones in the mixed space) by an mLE of approximately 0.04 (vertical dashed line in Fig.~\ref{figur6}).

\section{Machine learning}
\label{sec5:mach}

\subsection{Method}

In Sect.~\ref{sec4:res}, a correspondence between mLEs and $H$ was found. Both characteristic exponents were computed in a given point in the mixed space $\kappa\times x_0$, where $\kappa$ denotes $K$ or $k$. Hence, we are equipped with pairs of three-dimensional vectors in the form $(\kappa,x_0,{\rm mLE})$ and $(\kappa,x_0,H)$, where the first two components of each vectors are the same for a given map, i.e. the mLEs and $H$ were evaluated at the same points of the mixed space. A machine learning (ML) procedure is applied to those sets of vectors in order to predict $H$ values based only on the triples $(\kappa,x_0,{\rm mLE})$. The aim of this approach is (i) to train a classifier on $\approx 5000$ vector pairs from Sect.~\ref{sec4:res}, (ii) to compute a greater number of mLEs (i.e., in the mixed space with finer grid), resulting in $2\!-\!2.4\times 10^5$ mLEs, and (iii) to use the trained classifier to infer the $H$ values based on the mLEs. For this purpose, the nearest neighbor (NN) \cite{Altman,Barber,Cover,Hastie,Theodoridis} algorithm is employed, with an Euclidean metric to measure the distance. Its power lies in simplicity and accuracy. The training of an NN classifier is simply feeding it with the list of assignments \mbox{$\{(\kappa,x_0,{\rm mLE})_i\rightarrow H_i\}_{i=1}^{i=N_p}$}. Next, given a new mLE (not necessarily present in the training set) together with its location in the mixed space, its NN is found among the triples $(\kappa,x_0,{\rm mLE})$ that were used in the training process. The corresponding $H$ is ascribed to the new mLE. The ML is motivated by the tight correlation between the mLEs and $H$ found, exhibiting, however, some degree of scatter.\footnote{The described approach is implemented in \textsc{mathematica} via a built-in command \texttt{Predict} with \texttt{NearestNeighbors} chosen as the method of the ML.}

\subsection{Results}

\subsubsection{Chirikov standard map}
\label{sec51:res}

\begin{figure}
\centering
\includegraphics[width=\columnwidth]{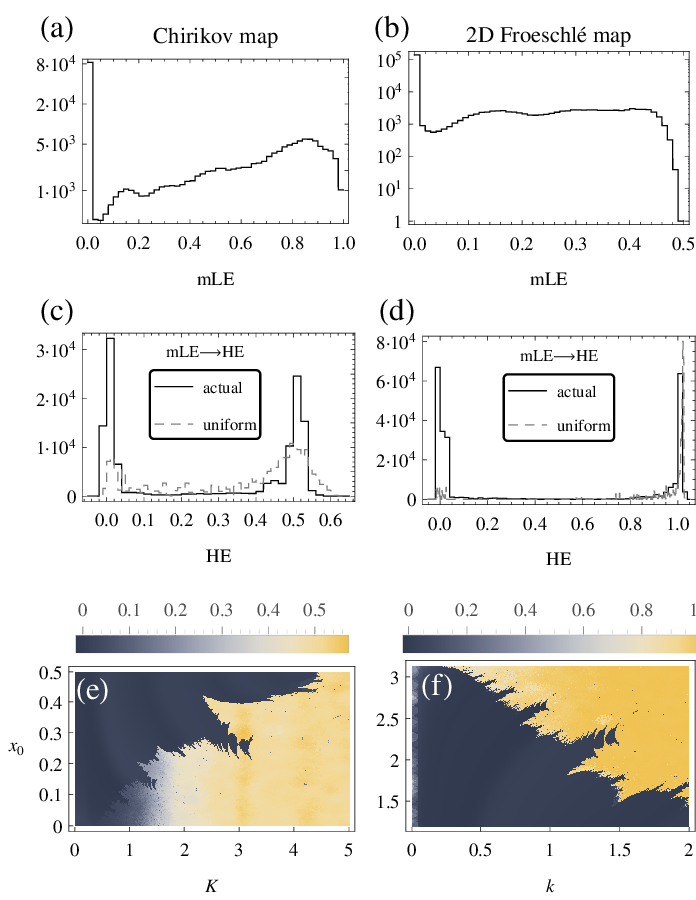}
\caption{{\it Left column}: Chirikov standard map. {\it Right column:} 2D Froeschl\'e map. (a)--(b) Actual mLE distributions for $\approx 2.0\cdot 10^5$ and $2.4\cdot 10^5$ values, respectively. (c)--(d) $H$ distributions obtained by the machine-learnt functions $p(x)$ with the NN method; solid black---actual distributions obtained by applying $p(x)$ to (a) and (b); dashed gray---comparison distributions obtained by applying $p(x)$ to artificial uniform distributions (at the level of $\approx 5000$). (e)--(f) $H$ distributions in the mixed space reproduced by applying appropriate $p(x)$ to (a) and (b). Note different color scales used.}
\label{figur7}
\end{figure}

ML training on a set of assignments $\{(K,x_0,{\rm mLE})_i\rightarrow H_i\}_{i=1}^{i=N_p}$ from Sect.~\ref{sec4:res} is performed. The output is a machine-learnt function, $p(x)$, whose input are the mLEs located in the mixed space, and its output is an estimate of $H$ corresponding to the same point. Next, $\approx 2.0\cdot 10^5$ mLEs are produced on a grid of $1001\times 201$ points in the mixed space $K\times x_0$. Resultant mLE distributions are shown in Fig.~\ref{figur7}(a) and (c), and are consistent with Fig.~\ref{figur2}(a). Two distinct peaks are easily distinguishable. The machine-learnt function $p(x)$ was then applied and the $H$ distribution reproduced with its aid is displayed in Fig.~\ref{figur7}(c). To emphasize that the underlying initial mLE distribution is crucial and $p(x)$ itself is not informative, the $H$ distribution is computed also for artificial, uniform mLE distributions. This resulted in uniformly mapped, via $p(x)$, $H$ distribution, overlaid with the actual one in Fig.~\ref{figur7}(c), significantly different from the real one.

These machine-learnt $H$ distributions were eventually mapped on the mixed space and the results are shown in Fig.~\ref{figur7}(e). Regular and chaotic regions are sharply divided, as in Fig.~\ref{figur1}, and the coverage appears to be nearly perfect. It is important to note that $p(x)$ is obtained only based on mLEs and $H$ both determinate, i.e., as described in Sect.~\ref{sec4:res}, extreme mLEs along the unstable line $x_0=0$ were discarded due to lack of a corresponding $H$. This is why the absolute color functions in Fig.~\ref{figur1}(b) and Fig.~\ref{figur7}(e) are different, yet the relative shapes of the distributions are apparently nearly identical.

\subsubsection{2D Froeschl\'e map}
\label{sec52:res}

The same was performed for the 2D Froeschl\'e map for $\approx 2.4\times 10^5$ mLEs on a grid of $501\times 486$ points in the mixed space of $k\times x_0$. The resulting mLE distribution is shown in Fig.~\ref{figur7}(b), and is consistent with Fig.~\ref{figur5}(a). The distribution in nearly uniform for non-zero mLEs, with a steep decrease just before 0.5, while in case of the Chirikov map, two peaks were easily distinguishable [compare with Fig.~\ref{figur2}(a)]. Next, the machine-learnt function $p(x)$ was was applied to the actual mLE distribution as well as to an artificial, uniform one, shown in Fig.~\ref{figur7}(d). The difference, between the two is most visible in the height of the peak near zero-values, as the rest of the real distribution is nearly uniform, hence not really that different from the artificial one. Eventually, this machine-learnt $H$ distribution was mapped on the mixed space and the result is displayed in Fig.~\ref{figur7}(f). Regular and chaotic regions are again sharply divided, as they were in Fig.~\ref{figur4}.

Finally, let us note that the mLE--HE relations, displayed as scatter plots in Fig.~\ref{figur3} and \ref{figur6}, have common features: a prominent plateau and a steep increase before. Left part of these relations is obviously influenced by the amount of nearly-zero mLEs, which is approximately linearly dependent on the size of the regular zone relative to the chaotic region. This means that $p(x)$ may be different if the mixed space is bounded differently, therefore the $H$ distributions obtained with the aid of $p(x)$ will have the relative heights of the peaks dependent on the ratio of regular and chaotic orbits in the part of the mixed space examined. Nevertheless, as the mLE and $H$ distributions in the chaotic zones are not entirely characterized by a single peak, the inferred $p(x)$ is likely to describe the mLE--HE relations correctly.

\section{Discussion and conclusions}
\label{sec6:disc}

It was suspected that chaotic time series might be generally characterized by $H\neq 1/2$. The mLE and $H$ distributions were computed for the 2D conservative Chirikov standard map \cite{Chirikov,Lichtenberg,Meiss}. Both characteristic exponents reveal remarkably similar structures in the mixed space of nonlinear parameter $K$ versus initial condition $x_0$ (see Fig.~\ref{figur1}). Moreover, a tight correlation between the $H$ estimates and mLEs was found, characterized by $\rho=0.83$. This is remarkable, as the two exponents are descriptors of different behavior: the mLE is a measure of sensitivity to initial conditions, interconnected with chaos, and $H$ is a measure of persistency (long-term memory, autocorrelation)---it characterizes the incremental trend in the data.

The investigated map yielded interesting results (Sect.~\ref{subsec31:res}), in particular: while regular motion corresponds to $H\sim 0$, the peak related to the chaotic zone is at $H=0.4-0.6$, and with a steady gradient in between (see Fig.~\ref{figur2}(a) and \ref{figur3}).

Motivated by the tight correlation between the mLE and $H$, an ML procedure, using the NN algorithm, was performed to reproduce the $H$ distribution based on the mLE distribution alone (Sect.~\ref{sec51:res}). Approximately 5000 points from the mixed space $K\times x_0$ were used for training, and then with $2\times 10^5$ mLEs, the $H$ values were retrieved. It should be emphasized that the shape of the mLE--HE relation is crucial here, as when an artificially, uniformly distributed set of mLEs was used, the retrieved $H$ distribution was different from the one computed directly from the time series [Fig.~\ref{figur7}(c)]. The NN procedure, applied to the true mLEs, allowed to reproduce the structure of the mixed space in great detail [Fig.~\ref{figur7}(e)].

To verify the robustness of the findings, the same analysis was performed for a 2D Froeschl\'e map. The general features of the mixed space $k\times x_0$ are similar to the Chirikov map's case (compare Fig.~\ref{figur1} and Fig.~\ref{figur4}), but the relation between mLE and $H$ (depicted in the scatter plot in Fig.~\ref{figur6}) is a bit different in character; the correlation is also rather tight ($\rho=0.75$), but with a very steep transition between low-$H$ (related to low mLEs) and high-$H$ values (related to high mLEs). Moreover, the scatter in $H$ decreases with increasing mLE. 

Using about 6000 points from the mixed space $k\times x_0$ together with the corresponding values of $H$ were used in the ML procedure, and next a more detailed structure of this mixed space was retrieved with $2.4\times 10^5$ mLEs [Fig.~\ref{figur7}(f)]. The discrepancy between the actual mLE distribution and an artificial uniform one was smaller than in case of the Chirikov standard map, as for non-zero mLEs both distributions were nearly uniform. The difference was manifested in the height of the peak related to low mLEs [Fig.~\ref{figur7}(b) and (d)].

To conclude, the important results obtained in this work are as follows:
\begin{enumerate}
\item For the low-dimensional conservative maps examined herein, i.e. the Chirikov standard map and the 2D Froeschl\'e map, the mLEs and $H$ estimates were found to be tightly correlated, and the correlations are positive.
\item The $H$ values corresponding to chaotic motion are systematically greater than the ones related to regular motion.
\item The chaotic zone is described by $H$ in the range $\sim 0.4-0.6$ for the Chirikov map, and $\sim 1$ for the 2D Froeschl\'e map; regular motion yields $H\sim 0$ for both maps.
\item Based on the correlation obtained, an ML was performed, and its results, applied to a much higher number of mLEs, allowed to reproduce the structure of the mixed space in great detail, including the bifurcating tongues.
\end{enumerate}

The HE appears to be an informative parameter that might find its place in the field of chaotic control, as it gives expectations about the general trend in the time series.
The cause underlying the mLE--HE relations, however, remains an open issue. Moreover, a question about its shape for different types of systems (non-symplectic, dissipative, higher dimensional, continuous, hyperchaotic, etc.) naturally arises. It can be expected that further exploration of this topic will lead to a deeper understanding of chaotic dynamical systems.

\section*{References}

\bibliography{mybibfile}

\end{document}